# CROSS-MODEL VERIFICATION OF WALL-BOUNDED FLOWS USING FINITE-JAX

**Arturo Rodriguez[1*], Avinash Potluri[1], Aryan Singh[2], Vyom Kumar[3], Kate Reza[4], Francisco O. Aguirre Ortega[1], Vineeth Vijaya Kumar[1], Noah L. Estrada[1]**

[1]Texas A&M University - Kingsville, Kingsville, TX 78363, USA

[2]Casady School, Oklahoma City, OK, USA

[3]Moreau Catholic High School, Hayward, CA 94544, USA

[4]University of Texas at El Paso, El Paso, TX 79968, USA

## ABSTRACT

Accurate prediction of wall-bounded flows remains central to advancing both theoretical understanding and computational methods in fluid mechanics. In this study, we perform a numerical simulation of channel flow using a complementary approach: a high-performance, differentiable finite-difference solver developed in JAX (Finite-JAX), and an analytical solution derived from the Navier-Stokes Equations, also known as the Hagen-Poiseuille equation. The solver is applied to the incompressible Navier-Stokes equations, along with appropriate boundary conditions, to capture canonical flow features, including velocity profiles and pressure gradients. Cross-model verification is conducted by systematically comparing numerical results between Finite-JAX and the analytical solution, with a focus on velocity distributions. In addition, numerical results are benchmarked against analytical solutions for the laminar regime, allowing direct quantification of the verification accuracy. Our findings demonstrate that cross-model verification not only strengthens confidence in simulation fidelity but also provides a pathway for integrating differentiable solvers with established computational fluid dynamics platforms, paving the way for future fluid flow research. The performance of Finite-JAX on Wall-Bounded Flows is 0.014765 in the L2 norm.

## NOMENCLATURE

| Variables | | | Subscripts and Superscripts | |
|---|---|---|---|---|
| $\vec{u}$ | velocity vector | m/s | $x$ | x-coordinate |
| $t$ | time | S | $y$ | y-coordinate |
| $\rho$ | density | kg/m$^3$ | ana | analytical solution |
| $p$ | pressure | Pa | $L2$ | L2 norm |
| $\nu$ | kinematic viscosity | $m^2/s$ | i | x-index |
| $F$ | forcing | | j | y-index |
| $u$ | velocity in the x-direction | m/s | n | time-index |
| $v$ | velocity in the y-direction | m/s | max | max value |
| $x$ | x-component of direction | m | | |
| $y$ | y-component of direction | m | | |
| $u$ | PDE solution of the x-component | | | |
| $\mu$ | dynamic viscosity | Pa·s | | |
| $H$ | y-direction of the channel/pipe | M | | |
| $L$ | loss | | | |
| $CFL$ | courant number | | | |
| $v$ | solution of the y-component | | | |

## 1. INTRODUCTION

The accuracy of continuous-wall flow modeling poses a challenge in fluid mechanics, with applications ranging from aerospace vehicle design and pipe flow transport to biomedical flows and energy systems. Despite the simplicity of canonical configurations, channel flow demonstrates the complexities of boundary-layer physics, including the interplay between viscous and inertial forces, the development of velocity gradients, and the dependence of flow behavior on the Reynolds number [1,2,3]. As such, it frequently serves as a validation case for numerical methods and the use of turbulence models [4].

In recent decades, advances in computational fluid dynamics (CFD) have significantly enhanced the accuracy of simulations of wall-bounded flows [5]. Commercial software, such as ANSYS Fluent, has been widely adopted in industry and academia due to its mature finite-volume method, robust solvers, and extensions to turbulence models [6]. In parallel, advancements in programming have introduced a new class of solvers, facilitated by platforms such as JAX, TensorFlow, and PyTorch, which are designed for high-performance computing and support machine learning workflows [7, 8]. These solvers are particularly attractive for modern applications involving uncertainty quantification, data assimilation, and physics-informed neural networks (PINNs) [9, 10, 11, 12].

The emergence of numerical solvers, such as Finite-JAX—a high-performance, finite-difference CFD code developed in JAX—offers a way to integrate traditional numerical methods with differentiable programming capabilities [13, 14, 15, 16, 17, 18]. However, the reliability of these frameworks first requires systematic verification and validation against established commercial solvers and analytical solutions. Model-to-model verification provides a structured approach to evaluating new high-fidelity tools, ensuring that predictions are consistent with those of validated CFD software across scenarios from a range of canonical flows.

In this work, we conduct a comprehensive study of channel flows across multiple models, verifying the results using Finite-JAX and the analytical solution. The solvers are applied to the incompressible Navier-Stokes equations with surface conditions that capture flow characteristics, including velocity distributions and pressure gradients. Results are compared with analytical solutions in the laminar regime, and the comparisons are extended to include friction factors and Reynolds-number scaling. Through this procedure, we can quantify the verification accuracy, establish confidence in the Finite-JAX differentiable CFD solver, and highlight the complementary roles of traditional and differentiable solvers in advancing the study of wall-bounded flows.

## 2. METHODS & EQUATIONS

We solve the incompressible Navier-Stokes equations in 2-D (channel flow with periodicity in the stream direction, no-slip walls above and below), with the streamwise constant a body force F (equivalent to a uniform pressure gradient):

$$\frac{\partial \vec{u}}{\partial t} + (\vec{u} \cdot \nabla)\vec{u} = -\frac{1}{\rho}\nabla p + \nu \nabla^2 \vec{u} + F\hat{e}_x \tag{1}$$

$$\nabla \cdot \vec{u} = 0 \tag{2}$$

Where $\vec{u} = (u, v)$, density $\rho$, kinematic viscosity $\nu$, and $F$ is a constant acceleration in the x-direction. In a fully developed laminar channel driven by a constant $-\partial p/\partial x$, the code sets $\partial p/\partial x = -\rho F$ so that the forcing term $F\hat{e}_x$ represents the pressure gradient.

### 2.1 Chorin Projection Operator Splitting Method

To enforce incompressibility at each step, the code uses a projection-splitting method introduced by Chorin, which uses the Poisson equation, which discretizes the source term b and assembles it with the following discretization:

$$\nabla^2 p = \rho \left[\frac{1}{\Delta t}\nabla \cdot \vec{u} - (\partial_x u)^2 - 2(\partial_y u)(\partial_x v) - (\partial_y v)^2\right] \tag{3}$$

On the grid (interior),

$$b_{j,i} = \rho\Big[\frac{1}{\Delta t}\left(\frac{u_{j,i+1} - u_{j,i-1}}{2\Delta x} + \frac{v_{j+1,i} - v_{j-1,i}}{2\Delta y}\right) - \left(\frac{u_{j,i+1} - u_{j,i-1}}{2\Delta x}\right)^2 - 2\left(\frac{u_{j+1,i} - u_{j-1,i}}{2\Delta y}\right)\left(\frac{v_{j,i+1} - v_{j,i-1}}{2\Delta x}\right) - \left(\frac{v_{j+1,i} - v_{j-1,i}}{2\Delta y}\right)^2\Big] \tag{4}$$

With this b, the pressure Poisson equation is discretised (five-point stencil) as:

$$\frac{p_{j,i+1} - 2p_{j,i} + p_{j,i-1}}{\Delta x^2} + \frac{p_{j+1,i} - 2p_{j,i} + p_{j-1,i}}{\Delta y^2} = b_{j,i} \tag{5}$$

Rearranged for Jacobi/Gauss-Seidel iteration:

$$p_{j,i} \leftarrow \frac{\left(p_{j,i+1} + p_{j,i-1}\right)\Delta y^2 + \left(p_{j+1,i} + p_{j-1,i}\right)\Delta x^2 - b_{j,i}\Delta x^2 \Delta y^2}{2(\Delta x^2 + \Delta y^2)} \tag{6}$$

Boundary conditions for p:

- Periodic in x: values at i = -1 and i = 0 wrap to i = $n_x - 2$ and i = $n_x - 1$ (implemented explicitly for the first/last columns).
- Neumann at walls $y = 0, H$: $\partial p/\partial y = 0 \rightarrow p_{0,i} = p_{1,i}, p_{n_y} - 1, i = \ p_{n_y-2,i}$.

**2.2 Velocity – Momentum Step**

After solving pressure, we proceed to calculate velocities explicitly using Euler's method, employing central differences for convection, the pressure gradient, and diffusion. For interior nodes:

$$\begin{aligned} u_{j,i}^{n+1} = u_{j,i}^{n} - u_{j,i}^{n}\frac{\Delta t}{\Delta x}\left(u_{j,i}^{n} - u_{j,i-1}^{n}\right) - v_{j,i}^{n}\frac{\Delta t}{\Delta y}\left(u_{j,i}^{n} - u_{j-1,i}^{n}\right) - \frac{\Delta t}{2\rho\Delta x}\left(p_{j,i+1} - p_{j,i-1}\right) \\ + \nu\Delta t\left[\frac{u_{j,i+1}^{n} - 2u_{j,i}^{n} + u_{j,i-1}^{n}}{\Delta x^2} + \frac{u_{j+1,i}^{n} - 2u_{j,i}^{n} + u_{j-1,i}^{n}}{\Delta y^2}\right] + F\,\Delta t \end{aligned} \tag{7}$$

$$\begin{aligned} v_{j,i}^{n+1} = v_{j,i}^{n} - u_{j,i}^{n}\frac{\Delta t}{\Delta x}\left(v_{j,i}^{n} - v_{j,i-1}^{n}\right) - v_{j,i}^{n}\frac{\Delta t}{\Delta y}\left(v_{j,i}^{n} - v_{j-1,i}^{n}\right) - \frac{\Delta t}{2\rho\Delta x}\left(p_{j,i+1} - p_{j,i-1}\right) \\ + \nu\Delta t\left[\frac{v_{j,i+1}^{n} - 2v_{j,i}^{n} + v_{j,i-1}^{n}}{\Delta x^2} + \frac{v_{j+1,i}^{n} - 2v_{j,i}^{n} + v_{j-1,i}^{n}}{\Delta y^2}\right] \end{aligned} \tag{8}$$

Boundary conditions for u:

- No-slip walls $y = 0, H: u = 0, v = 0\ at\ j = 0\ and\ j = n_y - 1$.
- Periodic in x: values at $i = 0\ and\ i = n_x - 1$ are updated using wrapped neighbors.

**2.3 Stopping Criterion**

$$udiff = \frac{\sum u^n - \sum u^{n-1}}{\sum u^n} < 10^{-6} \tag{9}$$

**2.4 Analytical Benchmark | Plane Poiseuille Profile**

With constant $-\partial p/\partial x = \rho F$ and no-slip at $y = 0, H$, the steady laminar solution is:

$$u_{ana}(y) = -\frac{1}{2\mu}\frac{\partial p}{\partial x}y(y - H) = -\frac{\rho F}{2\mu}y(y - H) \tag{10}$$

$$v_{ana} = 0 \tag{11}$$

The code compares the numerical solution $u(y)$ at a mid-plane $x = const$ to $u_{ana}(y)$, and reports the error:

$$L_2(u) = (\frac{1}{N}\sum_k [u_k - u_{ana}(y_k)]^2)^{1/2} \tag{12}$$

$$max|u - u_{ana}| \tag{13}$$

$$min|u - u_{ana}| \tag{14}$$

**2.5 Stability**

- Spatial operators are second-order accurate (central differences); time integration is first-order.
- Stability requires a suitable step obeying a CFL-like restriction combining convection and diffusion:

$$\Delta t \leq \min\left(\frac{CFL\Delta x}{|u|_{max}}, \frac{CFL\Delta y}{|v|_{max}}, \frac{CFL_v}{v}, \frac{\Delta x^2 \Delta y^2}{\Delta x^2 + \Delta y^2}\right) \tag{15}$$

With typical choices, CFL, $CFL_v \leq 0.5$ for robustness.

## 3. RESULTS AND DISCUSSION

Figure 1 shows the numerical solution of the streamwise velocity component, u (x, y), in the canonical two-dimensional channel flow. The color map displays the velocity magnitude, with the darkest color indicating low velocity near the wall, where the velocity is zero, and the lightest color representing the stream flow velocity. The distribution reveals a parabolic velocity profile, consistent with the classical pressure-driven Poiseuille flow solution. The velocity decreases near the walls due to the no-slip condition, increases monotonically through the shearing layers, and reaches a maximum velocity at the channel center.

What is particularly interesting in this figure is the invariance of the velocity contours along the flow direction. They are spatially uniform and reflect the fully developed flow state, in which streamwise gradients vanish, and the dynamics reduce to a balance between pressure forcing and viscous dissipation. The color bar quantifies the velocity magnitude, ranging from zero at the wall to the free stream at the center, which agrees with the theoretically derived analytical solution and the expectations of the forcing.

This result highlights the solver's ability to recover the fundamental benchmark of incompressible fluid mechanics. The parabolic structure here evidently serves not only to verify the numerical discretization and boundary conditions, but also to demonstrate that simple geometries capture the essence of viscous transport phenomena. Broadly speaking, this simulation serves as a step toward more complex scenarios—such as unsteady, turbulent, or non-Newtonian—where deviations from parabolic patterns enrich the Multiphysics scale of wall-bounded flows.

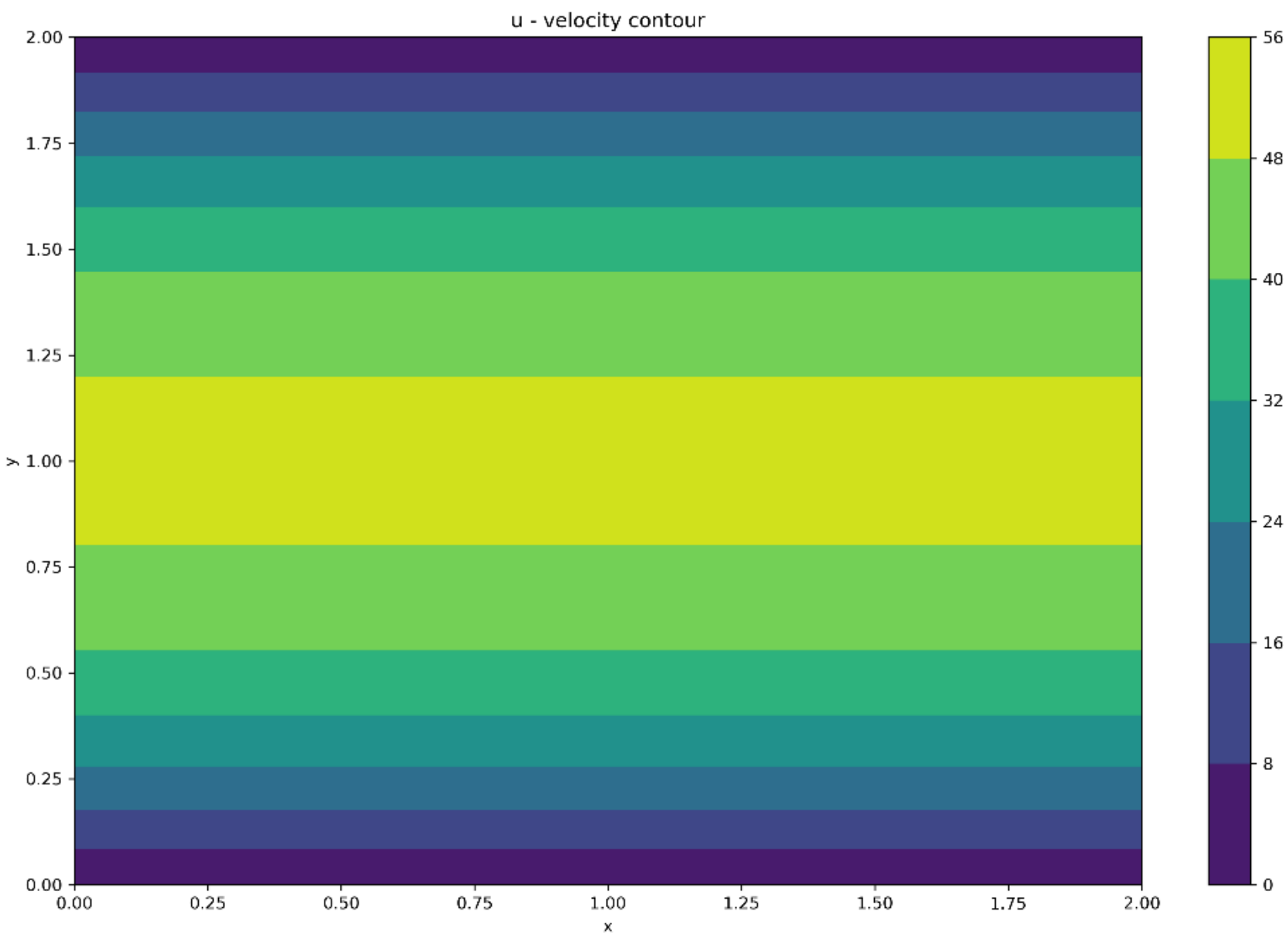

**Fig. 1.** Channel Flow Contour in m/s

**Table 1.** Error Norms

| L2 Norm | Max Norm | Min Norm |
|---|---|---|
| 1.4765e-02 m/s | 2.1140e-02 m/s | 0.0000e+00 m/s |

Figure 2 illustrates a direct comparison between the analytical solution of a developed Poiseuille flow and the numerical solution obtained by a JAX-based CFD solver. The horizontal axis represents the velocity component u, and the vertical axis represents the wall-normal coordinate y, extending from y = 0 to y = 2, corresponding to the channel height. The analytical profile consists of discrete points, whereas the numerical solution represents a continuous curve.

The parabolic shape of the velocity distribution is immediately apparent. The velocity vanishes at the surfaces, reflecting the no-slip condition, and increases to a maximum at the channel center. This demonstrates pressure-driven laminar channel flow, where the balance between viscous diffusion and the pressure gradient is evident, with a quadratic dependence on y. What is notable is the agreement between the two curves: the numerical solution obtained by coupling pressure and velocity reproduces the analytical results almost exactly, with a deviation that is imperceptible to the eye.

The agreement not only validates the correctness of the discretization and boundary conditions but also the implementation in the JAX solver, which precisely captures the pressure of the Poisson scheme and thereby preserves the incompressibility of fluid dynamics. This figure provides a strong benchmark; if the solver can replicate the canonical profile with high accuracy, one gains confidence in its application in more complex settings where analytical solutions do not exist. In computational fluid dynamics, verifying fundamental solutions is essential before extending numerical methods to the transition and turbulent regimes.

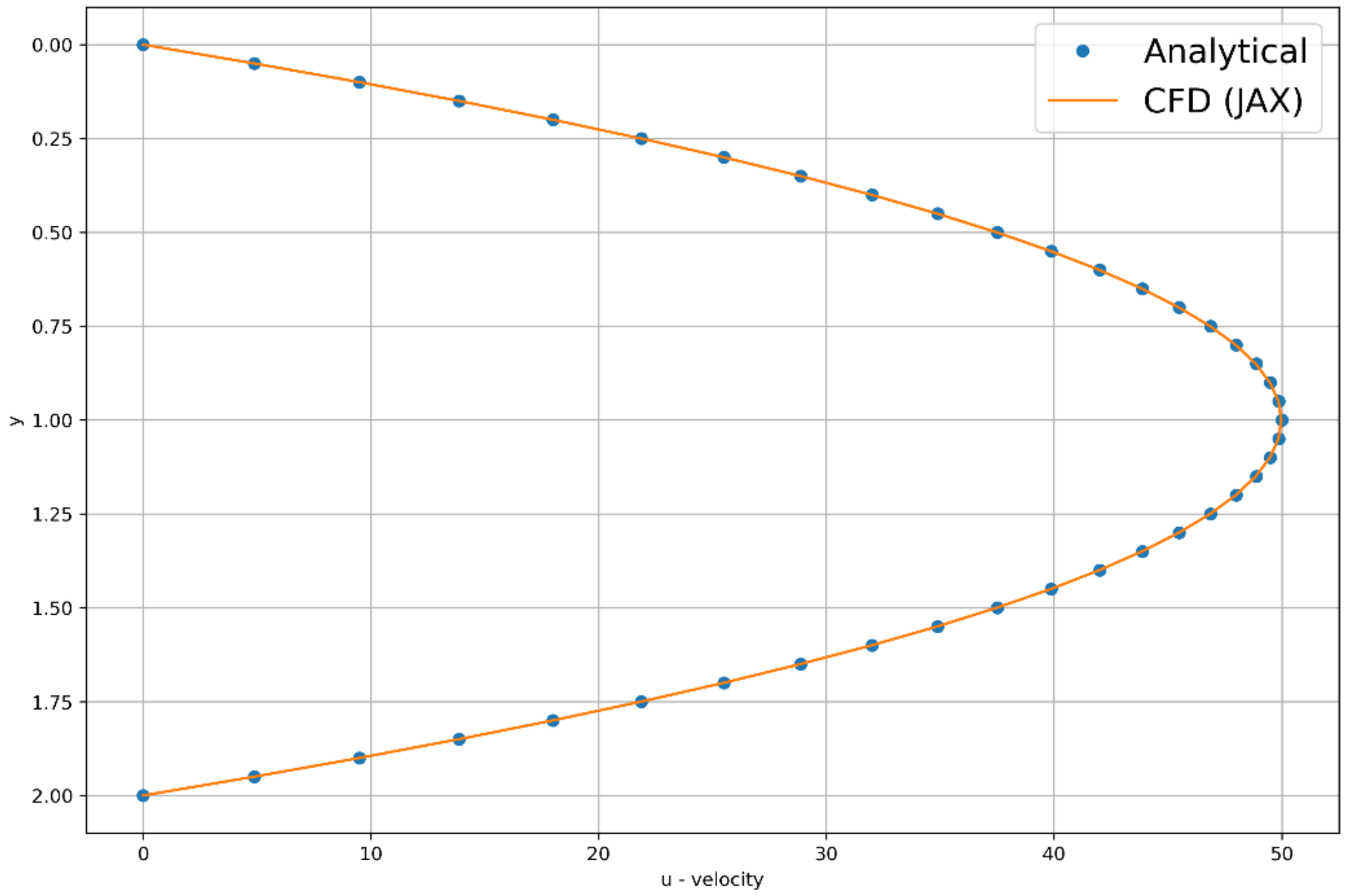

**Fig. 2.** Channel Flow Numerical vs. Analytical Solutions

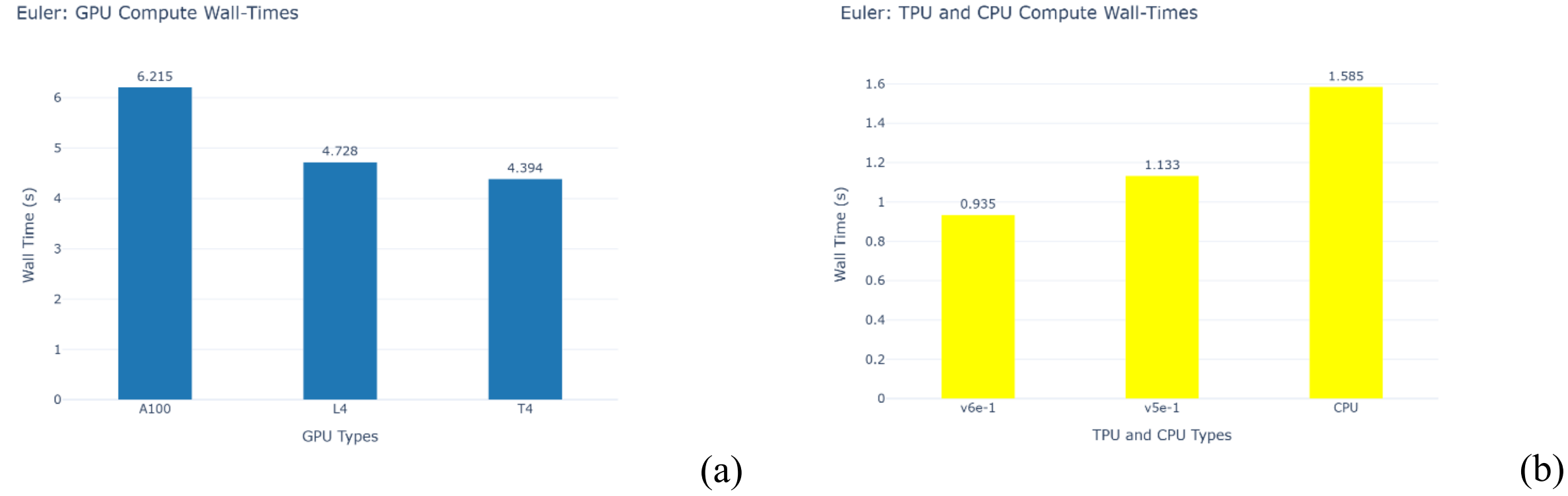

(a) (b)

**Fig. 3** (a) Euler: GPU Compute Wall-Times and (b) Euler: TPU and CPU Compute Wall-Times

Figure 3 presents a quantitative comparison of the wall times for computing the Euler discretization on different hardware accelerators. The left figure shows GPU-based performance, while the right shows TPU and CPU execution times. The vertical axis shows wall time in seconds, a measure of computational efficiency, and the horizontal axis identifies the hardware platform on which the code runs.

In the computing benchmarks, three architectures are examined: the A100, L4, and T4 GPUs. Somewhat counterintuitively, the A100, the flagship of GPU performance, takes the longest to compute, approximately 6.2 seconds, whereas the smaller T4 card yields faster results at approximately 4.4 seconds. The L4 takes approximately 4.7 seconds, which is in the middle of the two. The inversion of expectations demonstrates the interplay between algorithm structure, memory access patterns, and GPU architecture. This reflects the fact that capability does not translate directly into speedups, particularly in the structure of discrete Euler PDE solvers. These tests were performed using 64-bit double precision.

The TPU and CPU benchmarks reveal a complementary comparison. The v6e-1 TPU achieves the shortest computational time, approximately 0.935 seconds, which is less than the v5e-1 TPU (1.133 seconds) and the CPU (1.585 seconds). Here, hierarchical preemption is preserved: tensor processing is faster than general-purpose CPUs. The difference between v6e-1 and v5e-1 illustrates the gains in the hierarchy of successive generations of TPUs, offering reduced computation time for PDEs.

Together with the figures, this lesson is essential for scientific computing: performance is not only determined by peak FLOP counts, but also by hardware design, memory hierarchy, and algorithm structure. For PDE solvers, these results demonstrate the need for portable algorithms that efficiently adapt to heterogeneous architectures, thereby enhancing the fidelity and speed of flow and other simulation applications.

## 4. CONCLUSION

A two-dimensional incompressible Navier–Stokes solver based on the Chorin projection method was developed and validated for wall-bounded flows. The numerical formulation successfully captured velocity gradients in the boundary layer, with plane Poiseuille flow serving as a verification case. The stability limits imposed by the CFL condition were confirmed, and the efficiency of Jacobi and Gauss–Seidel iterations in solving the pressure Poisson equation was demonstrated.

Beyond classical CFD methods, this work highlights the advantages of integrating modern computational frameworks, such as JAX, into traditional solvers. These platforms enable scalable, differentiable simulations and provide a pathway toward machine learning-accelerated CFD. Future studies will extend the solver to three-dimensional turbulence, incorporate GPU-based acceleration, and develop hybrid ML–CFD strategies for large-scale wall-bounded flow applications.

## ACKNOWLEDGMENT

This project utilized Prof. Arturo Rodriguez's Startup funds, which Texas A&M University-Kingsville, the Department of Mechanical and Industrial Engineering, and the College of Engineering granted. This project was also supported by the DOE NNSA/MSIPP Grande CARES Consortium (GRANT13584020).

# APPENDIX

| ***Name*** | ***Symbol*** | ***Value*** | ***Units*** | ***Notes*** |
|---|---|---|---|---|
| Grid points in x | nx | 41 | - | Number of nodes along x |
| Grid points in y | ny | 41 | - | Number of nodes along y |
| Pressure Poisson Iterations | nit | 80 | - | Jacobi-like iterations per pressure solve |
| Domain Height | H | 2.0 | m | Domain spans |
| Grid spacing in x | dx | 0.05 | length | Uniform spacing |
| Grid spacing in y | dy | 0.05 | length | Uniform spacing |
| Density | rho | 2.0 | kg/m^3 | Fluid density |
| Kinematic viscosity | nu | 0.01 | m^2/s | Viscosity is used in momentum diffusion |
| Dynamic viscosity | mu | 0.02 | Pa-s | Appears in analytical solution |

| Body-force acceleration | F | 1.0 | m/s^2 | Constant x-direction forcing |
|---|---|---|---|---|
| Pressure gradient | dpdx | 2.0 | Pa/m | Equivalent constant dpdx driving the flow |
| Time step | dt | 0.1 | s | Explicit update step for u,v |
| Initial u field | u0 | zeros((ny,nx)) | - | Starts at rest |
| Initial v field | v0 | zeros((ny,nx)) | - | Starts at rest |
| Initial pressure | p0 | ones((ny,nx)) | - | Uniform initial guess |
| Convergence tolerance | tol | 1e-6 | - | Relative change threshold on sum(u) |
| Max iteration steps | max_steps | 100000 | - | Safety cap for while-loop convergence |
| JAX precision | jax_enable_x64 | True | - | Use float64 for better stability |
| JAX platform | jax_platform_name | cpu | - | Can be changed to “gpu” if available |